\documentclass[prl,aps,reprint,superscriptaddress, twocolumn]{revtex4-1}

\usepackage{graphicx,bm}
\usepackage{amsmath,amssymb}
\usepackage{epsfig}
\usepackage{epsf}
\usepackage[usenames]{color}

\usepackage{hyperref}
\hypersetup{colorlinks=true, linkcolor=blue, citecolor=red, urlcolor=blue}

\begin{document}

\title{Unidirectional magnetic coupling}

\author{H. Y. Yuan}

\affiliation{Institute for Theoretical Physics, Utrecht University, Princetonplein 5, 3584 CC Utrecht, The Netherlands}

\author{R. Lavrijsen}

\affiliation{Department of Applied Physics, Eindhoven University of Technology, P.O. Box 513, 5600 MB Eindhoven, The Netherlands}

\author{R. A. Duine}

\affiliation{Institute for Theoretical Physics, Utrecht University, Princetonplein 5, 3584 CC Utrecht, The Netherlands}

\affiliation{Department of Applied Physics, Eindhoven University of Technology, P.O. Box 513, 5600 MB Eindhoven, The Netherlands}

\date{\today}

\begin{abstract}
We show that interlayer Dzyaloshinskii-Moriya interaction in combination with non-local Gilbert damping gives rise to unidirectional magnetic coupling. That is, the coupling between two magnetic layers --- say the left and right layer --- is such that dynamics of the left layer leads to dynamics of the right layer, but not vice versa. We discuss the implications of this result for the magnetic susceptibility of a magnetic bilayer, electrically-actuated spin-current transmission, and unidirectional spin-wave packet generation and propagation. Our results may enable a route towards spin-current and spin-wave diodes and further pave the way to design spintronic devices via reservoir engineering.

\end{abstract}


\maketitle

\def\bD{{\bm D}}
\def\bx{{\bm x}}
\def\bk{{\bm k}}
\def\bK{{\bm K}}
\def\bq{{\bm q}}
\def\br{{\bm r}}
\def\bp{{\bm p}}
\def\bM{{\bm M}}
\def\bs{{\bm s}}
\def\bB{{\bm B}}
\def\bj{{\bm j}}
\def\bF{{\bm F}}
\def\id{{\rm d}}
\def\bE{{\bm E}}
\def\bmag{{\bm m}}
\def\bH{{\bm H}}

\def\br{{\bm r}}
\def\bv{{\bm v}}

\def\half{\frac{1}{2}}
\def\args{(\bm, t)}

{\it Introduction.} --- Non-reciprocal transmission of electrical signals lies at the heart of modern communication technologies. While semi-conductor diodes, as an example of an electronic component that underpins such non-reciprocity, have been a mature technology for several decades, new solutions are being actively pursued \cite{NE2020,Cheong2018}. Such research is spurred on by the emergence of quantum technologies that need to be read out electrically but should not receive unwanted back-action from their electronic environment.

Complementary to these developments, spintronics has sought to control electronic spin currents and, more recently, spin currents carried by spin waves --- i.e., magnons --- in magnetic insulators \cite{ChumakNP2015}. Devices that implement non-reciprocal spin-wave spin currents have been proposed \cite{Jamali2013,Lan2015,Grassi2020,Szulc2020}. Most of these proposals rely on dipolar interactions \cite{Damon1961,Camley1987,An2013,Kwon2016} or Dzyaloshinskii-Moriya interactions (DMI) \cite{MelcherPRL1973,Udvardi2009,MoonPRB2013,GarciaPRL2015,WangPRL2020}. Other proposals involve the coupling of the spin waves to additional excitations such that the spin waves are endowed with non-reciprocity. Examples are the coupling of the spin waves to magnetoelastic, optical, and microwave excitations \cite{WangPRL2019,YuPRL2019,TatenoPRAppl2020,Shah2020,ZhangPRAppl2020,Zhang2022}.

Most of these proposals have in common that they consider spin-wave dispersions that are asymmetric in wave vector. For example, due to the DMI spin waves at one particular frequency have different wave numbers and velocities for the two different directions. There are therefore spin waves travelling in both directions. This may be detrimental for some applications. For example, one would like to shield quantum-magnonic technologies from spin-current noise \cite{YuanPR2022}, and completely quench the spin-current transmission in one of the two directions along a wire.

Here, we propose a set-up that realizes unidirectional magnetic coupling between two magnetic layers or between two magnetic moments. The ingredients are DMI and dissipative coupling between the two layers or moments. The dissipative coupling takes the form of a non-local Gilbert damping and may arise, for example, from the combined action of spin pumping and spin transfer. Then, one magnet emits spin current when it precesses, which is absorbed by the other. The resulting dissipative coupling turns out to, for certain parameters, precisely cancel the DMI in one direction. As a result, an excitation of one of the magnets leads to magnetization dynamics of the other, but not vice versa. This yields spin-wave propagation that is truly uni-directional: for specific direction and magnitude of the external field, all spin waves travel in one direction only.

\begin{figure}
  \centering
  \includegraphics[width=0.45\textwidth]{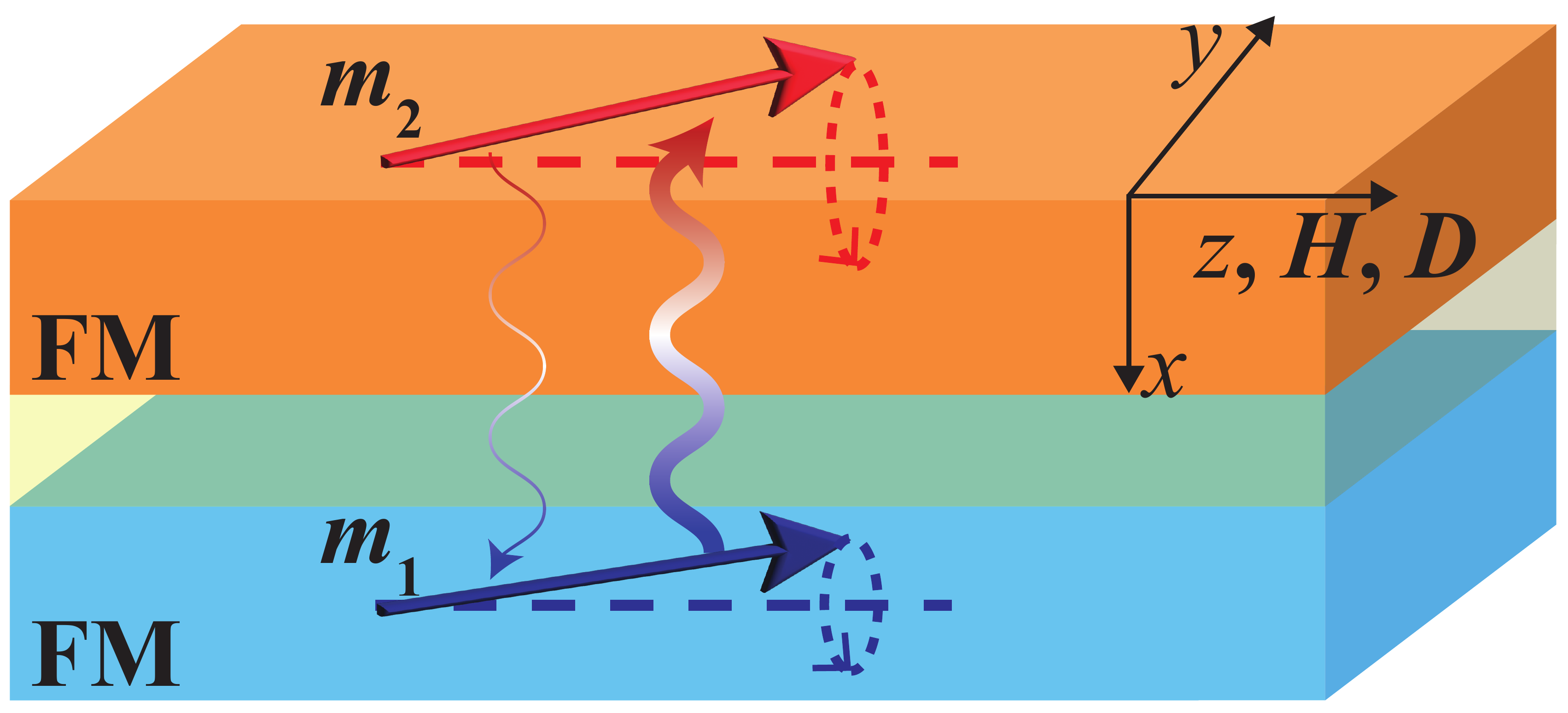}\\
  \caption{Schematic of two magnetic moments coupled by an interlayer DMI and by interlayer spin pumping. The dynamics of $\mathbf{m}_1$ induces the motion of $\mathbf{m}_2$, but not vice versa for appropriate parameters.}\label{fig1}
\end{figure}

{\it Minimal model.} --- Let us start with the minimal set-up that demonstrates the unidirectional coupling. We first consider two identical homogeneous magnetic layers that are coupled only by an interlayer DMI with Dzyaloshinskii vector $\bD$ and by interlayer spin pumping (see Fig. \ref{fig1}). The magnetization direction in the layers is denoted by $\bmag_i$, where $i \in \{1,2\}$ labels the two layers. We also include an external field $\bH$. The magnetic energy is given by
\begin{equation}
\label{eq:energybasic}
  E[\bmag_1,\bmag_2] = \bD \cdot \left( \bmag_1 \times \bmag_2 \right)- \mu_0 M_s \bH\cdot \left( \bmag_1+\bmag_2\right),
\end{equation}
where $M_s$ is the saturation magnetization of both layers and $\mu_0$ is the vacuum susceptibility. The magnetization dynamics of layer 1 is determined by the Landau-Lifshitz-Gilbert (LLG) equation
\begin{equation}
\frac{\partial \bmag_1}{\partial t} =  \frac{\gamma}{M_s}  \bmag_1 \times \frac{\delta E}{\delta \bmag_1} +  \alpha_{\rm nl} \bmag_1 \times \frac{\partial \bmag_2}{\partial t}~,
\end{equation}
where $\gamma$ is the gyromagnetic ratio and $\alpha_{\rm nl}$ characterizes the  strength of the non-local damping that in this set-up results from the combination of spin pumping and spin transfer torques, as described in the introduction. The equation of motion for the magnetization dynamics of the second layer is found by interchanging the labels 1 and 2 in the above equation. Working out the effective fields $\delta E/\delta \bmag_i$ yields
\begin{subequations} \label{eq_twolayers}
\begin{align}
\frac{\partial \bmag_1}{\partial t} &=  \frac{\gamma}{M_s}  \bmag_1 \times \left( \bmag_2 \times \bD - \mu_0  M_s \bH\right)  + \alpha_{\rm nl} \bmag_1 \times \frac{\partial \bmag_2}{\partial t},\\
\frac{\partial \bmag_2}{\partial t} &=  \frac{\gamma}{M_s}  \bmag_2 \times \left( \bD \times \bmag_1 -  \mu_0 M_s \bH\right)  + \alpha_{\rm nl} \bmag_2 \times \frac{\partial \bmag_1}{\partial t}\!,
\end{align}
\end{subequations}
where the sign difference in effective-field contribution from the DMI stems from the asymmetric nature of the DMI. We show now that depending on the magnitude and direction of the effective field, this sign difference leads for one of the layers to cancellation of the torques due to interlayer DMI and non-local damping. As the cancellation does not occur for the other layer, and because the DMI and non-local damping are the mechanisms that couple the layers in the model under consideration, this leads to uni-directional magnetic coupling.

Taking the external field to be much larger than the interlayer DMI, i.e., $\mu_0 |\bH| \gg |\bD|/M_s$, and taking $\alpha_{\rm nl} \ll 1$, we may replace $\partial \bmag_i/\partial t$ by $- \gamma \mu_0 \bmag_i \times \bH$ on the right-hand side of Eqs. \eqref{eq_twolayers} because the external field then is the dominant contribution to the precession frequency. For the field $\bH =\bD/\alpha_{\rm nl} \mu_0 M_s$, one then finds that
\begin{subequations}
\begin{align}
\frac{\partial \bmag_1}{\partial t} &= -\frac{\gamma}{\alpha_{\rm nl} M_s}  \bmag_1 \times \bD~,\label{llgm1}\\
\frac{\partial \bmag_2}{\partial t} &=   \frac{2 \gamma}{M_s}  \bmag_2 \times \left( \bD \times \bmag_1 \right)  -\frac{\gamma}{\alpha_{\rm nl} M_s}  \bmag_2 \times \bD ~. \label{llgm2}
\end{align}
\end{subequations}
Hence, the coupling between the two magnetic layers is unidirectional at the field $\bH =\bD/\alpha_{\rm nl} \mu_0 M_s$: the magnetization dynamics of layer 1 leads to dynamics of layer 2 as evidenced by Eq.~\eqref{llgm2}, but not vice versa as implied by Eq. \eqref{llgm1}. This one-way coupling is reversed by changing the direction of the field to $-\mathbf{H}$ or the sign of the non-local coupling $\alpha_{\rm nl}$.



\begin{figure}
  \centering
  \includegraphics[width=0.4\textwidth]{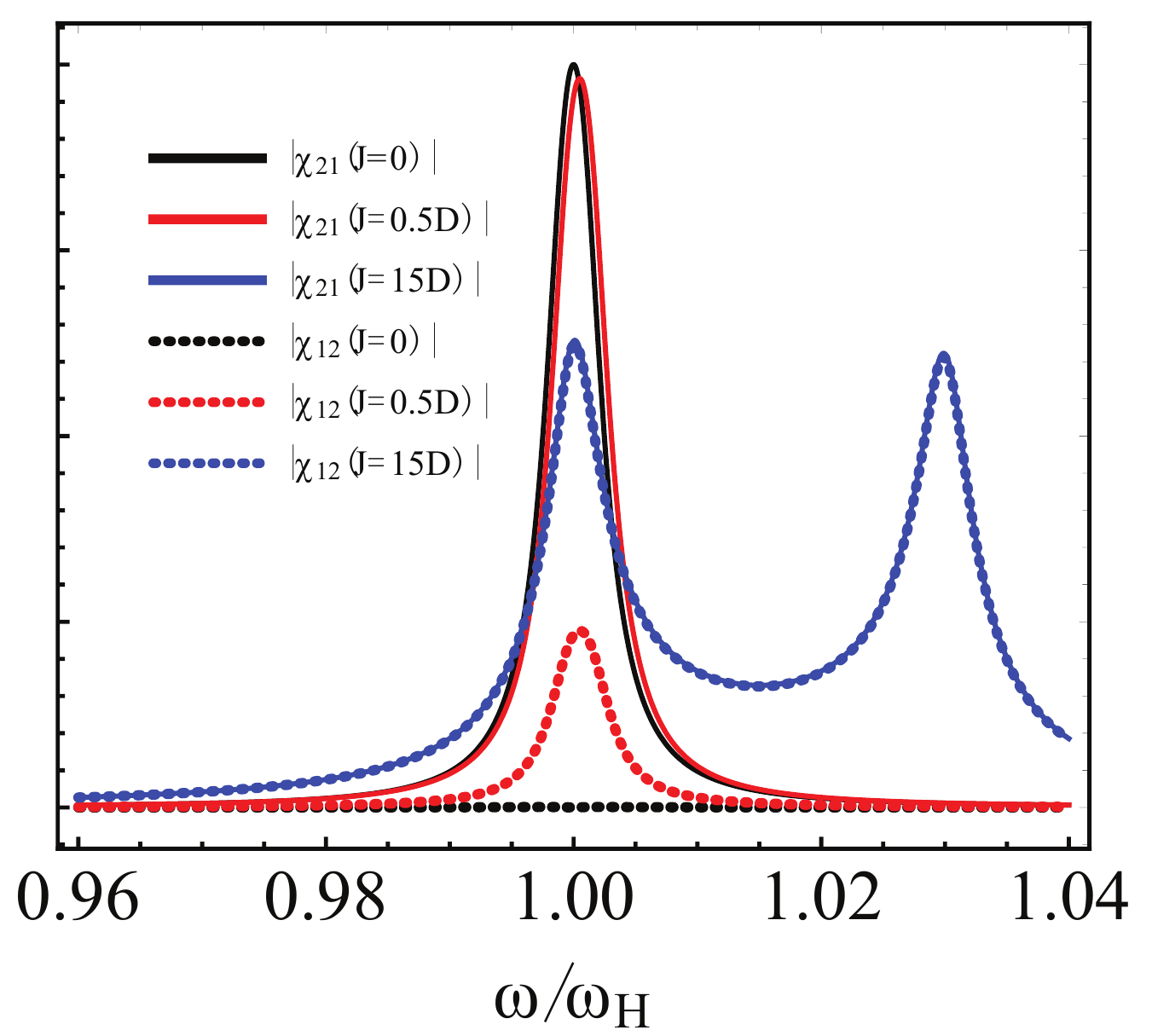}\\
  \caption{Magnetic susceptibilities of two magnetic layers as a function of frequency at different exchange couplings. $\omega_H \equiv \gamma (\mu_0 H + K/M_s)$. The resonance frequencies are located at the peak positions. The parameters are $D/\omega_H=0.001, \alpha_\mathrm{nl} =0.001, \alpha =0.002$.}\label{fig2}
\end{figure}

{\it Magnetic susceptibility.} --- Let us now take into account the Gilbert damping within the layers, exchange, and anisotropies and discuss the influence of the unidirectional coupling on the magnetic susceptibility. The energy now reads
\begin{eqnarray}
\label{eq:energyanisotropy}
&& E[\bmag_1,\bmag_2] = - J \bmag_1 \cdot \bmag_2 + \bD \cdot \left( \bmag_1 \times \bmag_2 \right)\nonumber \\&& - \mu_0 M_s \bH\cdot \left( \bmag_1+\bmag_2\right)  - \frac{K}{2} \left(m_{1,z}^2+m_{2,z}^2\right)~,
\end{eqnarray}
with the constant $K$ characterizing the strength of the anisotropy and $J$ the exchange. We shall focus on the ferromagnetic coupling ($J>0$) without loss of generality. The LLG equation now becomes
\begin{eqnarray}
\frac{\partial \bmag_1}{\partial t} &=&  \frac{\gamma}{M_s}  \bmag_1 \times \frac{\partial E}{\partial \bmag_1} + \alpha \bmag_1 \times \frac{\partial \bmag_1}{\partial t} \nonumber \\
&&+ \alpha_{\rm nl} \bmag_1 \times \frac{\partial \bmag_2}{\partial t}~,
\end{eqnarray}
with $\alpha$ the Gilbert damping constant of each layer, and where the equation for the second layer is obtained from the above by interchanging the labels $1$ and $2$. We take the external field in the same direction as the Dzyaloshinskii vector and $\bD = D \hat z$, $\bH = H \hat z$, while $\mu_0 M_s H,K \gg D$, so that the magnetic layers are aligned in the $\hat z$-direction. Linearizing the LLG equation around this direction we write $\bmag_i = (m_{i,x}, m_{i,y}, 1)^T$ and keep terms linear in $m_{i,x}$ and $m_{i,y}$. Writing $\phi_i = m_{i,x} - i m_{i,y}$, we find, after Fourier transforming to frequency space, that
\begin{eqnarray}
\chi^{-1} (\omega)\left(\begin{array}{c} \phi_1 (\omega) \\ \phi_2 (\omega)  \end{array} \right) =0~.
\end{eqnarray}
To avoid lengthy formulas, we give explicit results below for the case that $J=0$, while plotting the results for $J \neq 0$ in Fig. \ref{fig2}. The susceptibility tensor $\chi_{ij}$, or magnon Green's function, is given by
\begin{eqnarray}
 &&\chi (\omega) = \frac{1}{\left((1+i\alpha) \omega -\omega_0 \right)^2 - \left(\gamma D/M_s)^2-\alpha_{\rm nl}^2 \omega^2 \right)} \nonumber \\
  &&\times
 \left( \begin{array}{cc}
 (1+i\alpha) \omega -\omega_0 & i (\gamma D/M_s - \alpha_{\rm nl} \omega) \\
 -i (\gamma D/M_s + \alpha_{\rm nl} \omega) & (1+i\alpha) \omega -\omega_0
 \end{array} \right)~,
\end{eqnarray}
with $\omega_0 = \gamma (\mu_0 H + K/M_s)$ the ferromagnetic-resonance (FMR) frequency of an individual layer. The poles of the susceptibility determine the FMR frequencies of the coupled layers and are, for the typical case that $\alpha, \alpha_{\rm nl} \ll 1$, given by
\begin{equation} \label{eq:resonancefrequency}
\omega_{\pm} = \omega_{r,\pm} - i \alpha \omega_{r,\pm}~,
\end{equation}
with resonance frequency
\begin{equation}
\omega_{r,\pm} = \gamma (\mu_0 H + K/M_s \pm D/M_s)~.
\end{equation}
When $\gamma \mu_0 H = \left(1\mp \alpha_{\rm nl} \right) D/ ( \alpha_{\rm nl} M_s) -K/M_s \approx D/(\alpha_{\rm nl} M_s) -K/M_s $ we have for $J=0$ that $\chi_{12} (\omega_{r,\pm}) =0$ while $\chi_{21} (\omega_{r,\pm}) \neq 0$, signalling the non-reciprocal coupling. That is, the excitation of layer $1$ by FMR leads to response of magnetic layer $2$, while layer $1$ does not respond to the excitation of layer $2$. For opposite direction of field the coupling reverses: the excitation of layer $2$ by FMR leads in that case to response of magnetic layer $1$, while layer $2$ does not respond to the excitation of layer $1$. As is observed from Fig. \ref{fig2}, for finite but small $J \ll D$, the coupling is not purely unidirectional anymore but there is still a large non-reciprocity. For $J \gg D$, this non-reciprocity is washed out.

\begin{figure}
  \centering
  \includegraphics[width=0.45\textwidth]{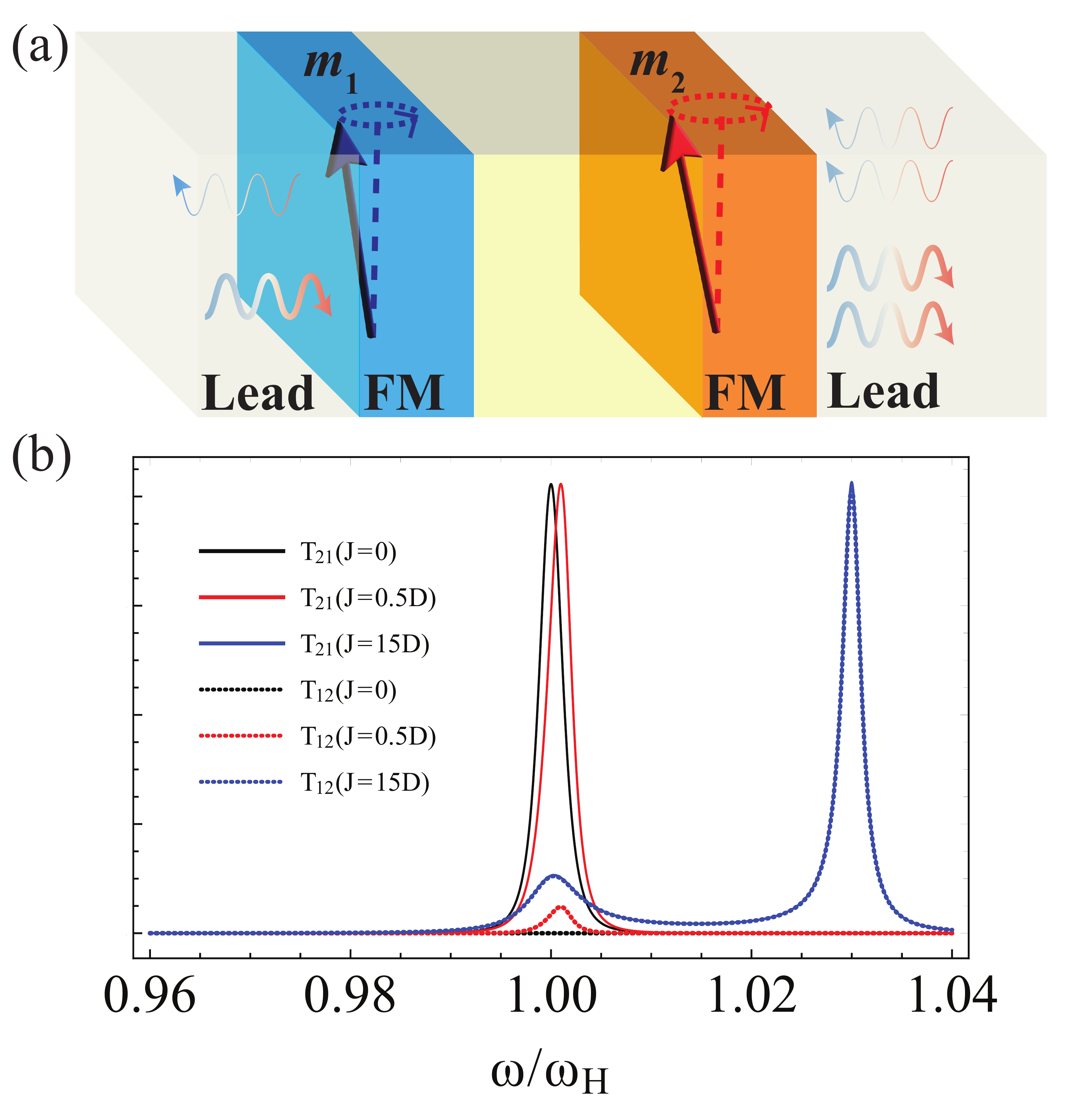}\\
  \caption{(a) Schematic of the system that the two coupled magnetic layers are sandwiched between two heavy-metal contacts. (b) Transmission of the hybrid system as a function of frequency.}\label{fig3}
\end{figure}

{\it Electrically-actuated spin-current transmission.} --- In practice, it may be challenging to excite the individual layers independently with magnetic fields, which would be required to probe the susceptibility that is determined above. The two layers may be more easily probed independently by spin-current injection/extraction from adjacent contacts. Therefore, we consider the situation that the two coupled magnetic layers are sandwiched between heavy-metal contacts (see Fig. \ref{fig3}(a)). In this set-up, spin current may be transmitted between the two contacts through the magnetic layers.


Following the Green's function formalism developed by Zheng {\it et al.} \cite{ZhengPRB2017}, the spin-current from the left (right) lead to its adjacent magnetic layer is determined by the transmission function of the hybrid system $T_{12}$ ($T_{21}$) given by
\begin{equation}
\label{eq:transmission}
 T_{ij} (\omega)= {\rm Tr} \left[ \Gamma_i (\omega) G^{(+)} (\omega) \Gamma_j (\omega) G^{(-)}(\omega)\right]~.
\end{equation}
Here, $G^{(+)}(\omega)$ is the retarded Green's function for magnons in contact with the metallic leads that is determined by Dyson's equation $\left[G^{(+)}\right]^{-1}(\omega) = \chi^{-1} (\omega) - \Sigma^{(+)}_1(\omega) - \Sigma^{(+)}_2 (\omega)$, where the retarded self energy $\hbar \Sigma^{(+)}_i (\omega)$ accounts for the contact with the metallic lead $i$. These self energies are given by
\begin{eqnarray}
 \hbar \Sigma^{(+)}_1 (\omega) =  - i  \hbar \alpha'_1 \left(\begin{array}{cc}
     \omega & 0\\
     0 & 0 \end{array}
  \right)~,
\end{eqnarray}
and
\begin{eqnarray}
\hbar \Sigma^{(+)}_2 (\omega) =  - i \hbar  \alpha'_2 \left(\begin{array}{cc}
0 & 0\\
0 & \omega \end{array}
\right)~.
\end{eqnarray}
The rates for spin-current transmission from the heavy metal adjacent to the magnet $i$ into it, are given by $\Gamma_i (\omega) = -2 {\rm Im} \left[ \Sigma_i^{(+)} (\omega) \right]/\hbar$. The couplings $\alpha'_i = \gamma Re[g^{\uparrow\downarrow}_i]/4 \pi M_s d_i$ are proportional to the real part of the spin-mixing conductance per area $g^{\uparrow\downarrow}_i$ between the heavy metal and the magnetic layer $i$, and further depend on the thickness $d_i$ of the magnetic layers. Finally, the advanced Green's function is $G^{(-)} (\omega) = \left[G^{(+)}\right]^\dagger$.

In the analytical results below, we again restrict ourselves to the case that $J=0$ for brevity, leaving the case $J \neq 0$ to plots. Using the above ingredients, Eq.~(\ref{eq:transmission}) is evaluated. Taking identical contacts so that $\alpha'_1=\alpha'_2\equiv \alpha'$, we find that
\begin{equation}
 T_{12} = \frac{4 (\alpha')^2 \omega^2 \left(\gamma D/M_s + \alpha_{\rm nl} \omega\right)^2}{|C(\omega)|^2}~,
\end{equation}
while
\begin{equation}
T_{21} = \frac{4 (\alpha')^2 \omega^2 \left(\gamma D/M_s - \alpha_{\rm nl} \omega\right)^2}{|C(\omega)|^2}~,
\end{equation}
with
\begin{eqnarray}
&& C (\omega) = \left[\omega_H- (1 +i (\alpha -\alpha_{\rm nl} +\alpha') )\omega \right] \cdot
        \nonumber \\ &&     \left[\omega_H - (1 +i (\alpha + \alpha_{\rm nl} +\alpha') )\omega \right] -(\gamma D/M_s)^2\!\!\!\!\! ~~~.
\end{eqnarray}
From the expression for $C(\omega)$ it is clear that, since $\alpha, \alpha_{\rm nl}, \alpha' \ll 1$, the transmission predominantly occurs for frequencies equal to the  resonance frequencies $\omega_{r,\pm}$ from Eq. \eqref{eq:resonancefrequency}. Similar to the discussion of the susceptibilities, we have for fields $\gamma \mu_0 H =  D/ \alpha_{\rm nl} -K/M_s$ that the transmission $T_{12} (\omega = D/\alpha_{\rm nl}) \neq 0$, while $T_{21} (\omega = D/\alpha_{\rm nl}) = 0$. As a result, the spin-current transmission is unidirectional at these fields. For the linear spin-conductances $G_{ij}$, given by $G_{ij} = \int \hbar \omega (-N' (\hbar \omega)) T_{ij} (\omega)$, we also have that $G_{12} \neq 0$, while $G_{21} =0$. Here, $N(\hbar \omega) = [e^{\hbar \omega/k_B T}-1]^{-1}$ is the Bose-Einstein distribution function at thermal energy $k_BT$. For the opposite direction of external field we have $G_{12} = 0$, while $G_{21} \neq 0$.
Like in the case of the susceptibility discussed in the previous section, a finite but small exchange coupling makes the spin current transport no longer purely unidirectional, while maintaining a large non-reciprocity (see Fig. \ref{fig3}(b)).


{\it Spin-wave propagation.} --- Besides the unidirectional coupling of two magnetic layers, the above results may be generalized to a magnetic multilayer, or, equivalently, an array of coupled magnetic moments that are labeled by the index $i$ such that the magnetization direction of the $i$-th layer is $\bmag_i$. This extension allows us to engineer unidirectional spin-wave propagation as we shall see below. We consider the magnetic energy
\begin{eqnarray}
\label{eq:energybasicmultilayer}
E[\bmag] &=& \sum_k \left[ \bD \cdot \left( \bmag_k \times \bmag_{k+1} \right)  - \mu_0 M_s \bH\cdot \bmag_k \right]~,
\end{eqnarray}
and find --- within the same approximations as for our toy model above --- for the magnetization dynamics that
\begin{equation}
\label{eq:llgmultilayer}
  \frac{\partial \bmag_k}{\partial t} =   \frac{2 \gamma}{M_s}  \bmag_k \times \left( \bD \times \bmag_{k-1} \right)  -\frac{\gamma}{\alpha_{\rm nl} M_s}  \bmag_k \times \bD~,
\end{equation}
for the field $\bH =\bD/\alpha_{\rm nl} \mu_0 M_s$. This shows that for these fields the magnetic excitations travel to the right --- corresponding to increasing index $k$ --- only. The direction of this one-way propagation is reversed by changing the magnetic field to $-\mathbf{H}$ or by changing the sign of the nonlocal damping.

To study how spin waves propagate in an array of coupled magnetic moments described by the Hamiltonian in Eq.~(\ref{eq:energybasicmultilayer}). We start from the ground state $\mathbf{m}_k=(0,0,1)^T$ and perturb the left-most spin ($k=0$) to excite spin waves. Since the dynamics of this spin is not influenced by the other spins for the field $H =D/ \alpha_{\rm nl} \mu_0 M_s$, its small-amplitude oscillation can be immediately solved as $\phi_0(t) = \phi_0(t=0) \exp(-i\omega_0t-\alpha \omega_0t)$ with $\phi_k=m_{k,x}-im_{k,y}$ as used previously. The dynamics of the spins to the right of this left-most spin is derived by solving the LLG equation (\ref{eq:llgmultilayer}) iteratively, which yields
\begin{equation}
\label{eq:spatialsw_genenral}
\phi_k(t) = \phi_0(t=0)e^{-i\omega_0t}\frac{e^{-\alpha \omega_0 t}}{k!}(-2 \alpha_{\rm nl} \omega_0 t)^k~,
\end{equation}
where $k=0,1,2,...N-1$.

To guarantee the stability of the magnetization dynamics, the dissipation matrix of the $N$-spin system should be negative-definite, which imposes a constraint on the relative strength of Gilbert damping and non-local damping, i.e., $\alpha>2\alpha_{\rm nl} \cos \frac{\pi}{N+1}$. For an infinitely-long chain $N \rightarrow \infty$, we have $\alpha>2\alpha_{\rm nl}$. Physically, this means that the local dissipation of a spin has to be strong enough to dissipate the spin current pumped by its two neighbors. For a spin chain with finite number of spins, $\alpha=2|\alpha_{\rm nl}|$ is always sufficient to guarantee the stability of the system. Taking this strength of dissipation simplifies Eq.~(\ref{eq:spatialsw_genenral}) to
\begin{equation}
\phi_k(t) = \phi_0(t=0)e^{-it/(\alpha \tau)}\frac{e^{-t/\tau}}{k!}(-t/\tau)^k~,
\end{equation}
where $\tau^{-1}=\alpha \omega_0$ is the inverse lifetime of the FMR mode. This spatial-temporal profile of spins is the same as a Poisson distribution with both mean and variance equal to $\sigma=t/\tau$ except for a phase modulation, and it can be further approximated as a Gaussian wavepacket on the time scale $t\gg \tau$, i.e.
\begin{equation}
\phi (x) = \frac{\phi_0(t=0)e^{-it/(\alpha \tau)}}{\sqrt{2\pi\sigma}} e^{-\frac{(x-\sigma)^2}{2\sigma}}.
\end{equation}
Such similarity suggests that any local excitation of the left-most spin will generate a Gaussian wavepacket propagating along the spin chain. The group velocity of the moving wavepacket is $v=a/\tau$, where $a$ is the distance between the two neighboring magnetic moments. The width of the wavepacket spreads with time as $a\sqrt{t/\tau}$, which resembles the behavior of a diffusive particle. After sufficiently long time, the wavepacket will collapse.

On the other hand, the excitation is localized and cannot propagate when the right-most spin ($k=N-1$) is excited, because its left neighbor, being in the ground state, has zero influence on its evolution. These results demonstrate the unidirectional properties of spin-wave transport in our magnetic array.

{\it Discussion, conclusion, and outlook.} --- We have shown that the ingredients for unidirectional coupling between magnetic layers or moments are that they are coupled only by DMI and non-local Gilbert damping. While in practice it may be hard to eliminate other couplings, the DMI and non-local coupling need to be sufficiently larger than the other couplings to observe unidirectional coupling.

There are several systems that may realize the unidirectional coupling we propose. A first example is that of two magnetic layers that are coupled by a metallic spacer. Such a spacer would accommodate non-local coupling via spin pumping and spin transfer. For a spacer that is much thinner than the spin relaxation length, we find, following Refs.~\cite{TserkovnyakRMP2005,HeinrichPRL2003,WoltersdorfPRL2007}, that $ \alpha_{\rm nl} = \gamma \hbar Re[\tilde  g^{\uparrow\downarrow}]/4 \pi  d M_s$, with $\tilde g^{\uparrow\downarrow}$ the spin-mixing conductance of the interface between the magnetic layers and the spacer, $d$ the thickness of the magnetic layers. For simplicity, we took the magnetic layers to have equal properties. The two magnetic layers may be coupled by the recently-discovered interlayer DMI \cite{Fernandez2019, Han2019}, tuning to a point (as a function of thickness of the spacer) where the ordinary RKKY exchange coupling is small.
We estimate $\alpha_{\rm nl} = 4.5 \times 10^{-3}$ for $d=20$ nm, $Re[\tilde  g^{\uparrow\downarrow}]=4.56 \times 10^{14} ~\mathrm{\Omega^{-1} m^{-2}}$ and $M_s = 1.92 \times 10^5~\mathrm{A/m}$ (YIG$|$Pt). The required magnetic field for unidirectional magnetic coupling is then around 4.5 T for $D=1$ mT.
Another possible platform for realizing the unidirectional coupling is the system of Fe atoms on top of a Pt substrate that was demonstrated recently \cite{Steinbrecher2019}. Here, the relative strength of the DMI and exchange is tuned by the interatomic distance between the Fe atoms. Though not demonstrated in this experiment, the Pt will mediate non-local coupling between the atoms as well. Hence, this system may demonstrate the unidirectional coupling that we proposed.

The non-local damping is expected to be generically present in any magnetic material and does not require special tuning, though it may be hard to determine its strength experimentally. Hence, an attractive implementation of the unidirectional coupling would be a magnetic material with spins that are coupled only via DMI, without exchange interactions.  While such a material has to the best of our knowledge not been discovered yet, it is realized transiently in experiments with ultrafast laser pulses \cite{Kerber2020}. Moreover, it has been predicted that high-frequency laser fields may be used to manipulate DMI and exchange, even to the point that the former is nonzero while the latter is zero \cite{Stepanov2017, Juan2019}.

Possible applications of our results are spin-wave and spin-current diodes and magnetic sensors, where a weak field signal can be amplified and transported through the unidirectional coupling to the remote site to be read out without unwanted back-action.
Finally, we remark that the unidirectional magnetic coupling that we propose here may be thought of as reservoir engineering, cf. Ref.~\cite{MetelmannPRX2015}. In our proposal, the reservoir is made up by the degrees of freedom that give rise to the non-local damping, usually the electrons. We hope that this perspective may pave the way for further reservoir-engineered magnetic systems

{\it Acknowledgements.} --- It is a great pleasure to thank Mathias Kl\"aui and Thomas Kools for discussions. H.Y.Y acknowledges the European Union's Horizon 2020 research and innovation programme under Marie Sk{\l}odowska-Curie Grant Agreement SPINCAT No. 101018193. R.A.D. is member of the D-ITP consortium that is funded by the Dutch Ministry of Education, Culture and Science (OCW). R.A.D. has received funding from the European Research Council (ERC) under the European Union's Horizon 2020 research and innovation programme (Grant No. 725509). This work is in part funded by the project  ``Black holes on a chip" with project number OCENW.KLEIN.502 which is financed by the Dutch Research Council (NWO).

\end{document}